\documentclass[letterpaper]{svjour2}
\smartqed
\newcommand{\ve}[1]{{\rm{\bf {#1}}}}
\newcommand{\syn}{^{\rm syn}}
\newcommand{\obs}{^{\rm obs}}

\newcommand{\mg}{^{\rm mg}}
\newcommand{\nm}{^{\rm nm}}
\newcommand{\htt}{\hat{t}}
\renewcommand{\thetable}{\Roman{table}}
\usepackage{graphicx}
\usepackage{amssymb}
\usepackage{amsmath}
\begin{document}

\title{MAGNETIC FIELD VECTOR RETRIEVAL WITH THE HELIOSEISMIC
AND MAGNETIC IMAGER}

\author{J.M.~BORRERO \and S.~TOMCZYK \and A.~NORTON \and T.~DARNELL \and J.~SCHOU
 \and P.~SCHERRER \and R.~BUSH \and Y.~LIU}

\institute{J.M.~ Borrero, S.~Tomczyk, A.~Norton$^{*}$, T.~Darnell \at
                 High Altitude Observatory, National Corporation for \\
                 Atmospheric Research, 3450 Mitchell Lane, Boulder CO 80301, USA\\
                 \email{borrero@ucar.edu, tomczyk@ucar.edu, norton@ucar.edu,
		 tdarnell@ucar.edu}\\
		 \and
           J.~Schou, P.~Scherrer, R.~Bush, Y.~Liu \at
	         Hansen Experimental Physics Laboratory, Stanford University \\
		 Stanford, CA 94305, USA\\
		 \email{schou@stanford.edu, pscherrer@stanford.edu, 
		 rock@stanford.edu, yliu@stanford.edu}\\
                 \and
           $^{*}$ Current address: National Solar
                  Observatory, 950 N Cherry Avenue, Tucson AZ 85719}

\date{Received: 8 May 2006 / Accepted: 11 November 2006 }
\maketitle
\begin{abstract}
We investigate the accuracy to which we can retrieve
the solar photospheric magnetic field vector using the
Helioseismic and Magnetic Imager (HMI) that will fly onboard
of the Solar Dynamics Observatory (SDO) by inverting simulated 
HMI profiles. The simulated profiles realistically take into account the effects
of the photon noise, limited spectral resolution, instrumental 
polarization modulation, solar {\it p} modes and temporal averaging.
The accuracy of the determination of the magnetic field
vector is studied considering the different operational modes 
of the instrument.
\end{abstract}
%
\section{Introduction}%

The Solar Dynamics Observatory is scheduled to launch from Cape
Canaveral in September 2008 on an Atlas V Booster.
Onboard this satellite there will be several instruments dedicated
to the study of the solar photospheric and coronal magnetic fields 
and their relation to the interplanetary medium, space weather and 
terrestrial climatology. The Helioseismic and Magnetic Imager is
an instrument developed by a collaboration between Stanford University, Lockheed-Martin 
Solar and Astrophysics Laboratory and the High Altitude Observatory.
HMI consists of a combination of a Lyot filter and two Michelson 
interferometers that will measure the full Stokes vector 
$\ve{I} = (I, Q, U, V)$ at 
several wavelengths across the Fe {\sc I} 6173.3 \AA~line.
HMI will have twin CCD cameras that will operate independently. The first
one, to be referred to as the ``Doppler camera'', will be devoted to the measurement 
of the line-of-sight component of the magnetic field and velocity, like MDI 
on SOHO. The second camera, hereafter referred to as the ``Vector Camera'', 
will measure the vector magnetic field and line-of-sight velocities. Each CCD consists of 
a 4096 $\times$ 4096  pixel array with a pixel size of 0.5 arcsec. Together with SDO's
geosynchronous orbit, this will allow for a near continuous monitoring of
the solar (full disk) vector magnetic field.

In this paper we study the accuracy to which we can retrieve the
magnetic field vector and other physical parameters of the solar photosphere
using the data from the HMI vector camera. Details about how the vector camera
records the data are presented in Section 2. In Section 3 we describe the
main sources of uncertainty in the data (photon noise, limited spectral resolution,
solar oscillatory modes) and how we simulate HMI data by taking into account those
effects. Once we account for those processes, we are able to use a Stokes synthesis code
to compute polarization profiles of the Fe {\sc I} 6173.3 \AA~line as if they were observed
by HMI. A large database of profiles is built by using an array
of atmospheric parameters. Using a Stokes inversion code (described in Section 4)
the physical parameters used in the synthesis can be recovered.
The errors in the retrieval depending upon the choice of the polarization modulation 
are studied in Section 5. The effects of the photon noise are considered in Section 6. 
Section 7 investigates the improvement in the retrieval
when the data are averaged in time. Our conclusions are finally given in Section 8.

\section{HMI Operational Modes}%

%
\paragraph{Line selection:}
The Helioseismic and Magnetic Imager will obtain two-dimensional images of the full solar disk
at several wavelengths across the neutral iron line at 6173.3 \AA. This
spectral line was chosen after a careful comparison with other candidates: Ni {\sc I} 6767.8 and 
Fe {\sc I} 6302.5 (see Norton {\it et al}. 2006). This line possesses 
a Land\'e factor of g$_{eff}=$2.5 and therefore is more suitable for the investigation 
of the magnetic field than its MDI counterpart Ni {\sc I} (g$_{eff}=$1.5) 6767.8 \AA~
(see Scherrer {\it et al}. 1995). In addition, its large slope in the wings ensures a high
accuracy in the derived velocity signals (see Cabrera Solana, Bellot Rubio, and del Toro Iniesta, 
2005). It is also relatively free of blends which makes this line very convenient 
for HMI purposes, since the geosynchronous orbit of SDO will induce Doppler 
shifts as large as $\pm 4$ km s$^{-1}$ due to spacecraft velocities.

\paragraph{Filter profiles:}
The combined Lyot-Michelson filter system in HMI produces a transmission profile with a FWHM
of 76 m\AA. Tuning positions are separated by 69 m\AA. Although there are a number
of different configurations and possible tuning positions available, in this work, we will
restrict ourselves to studying a configuration in which the spectral line is scanned across five
wavelength positions with the middle one located at the central laboratory wavelength $\lambda_0$
of the spectral line. Figure 1 displays the filter profiles at each wavelength position. In the
remainder of the paper we will denote each of the filter profiles as: $\mathcal{G}_i(\lambda)$, where 
$i=1,...,M=5$.

\paragraph{Filter sequence and polarization modulation:}
For each tuning position ($\lambda_i$) and time ($t_n$), HMI will measure
a linear combination [$\mathcal{O}(\lambda_i,t_n)$]
of the components of the solar Stokes vector\footnote{In compact notation, the
components of the Stokes vector are expressed as $I_k$ ($k=1,...4$) with: 
$I_1=1$, $I_2=Q$, $I_3=U$, $I_4=V$}:

\begin{eqnarray}
\mathcal{O}_p(\lambda_i, t_n) & = & \sum_{k=1}^{4} \mathcal{M}_{pk} \widetilde{I}\obs_k 
(\lambda_i,t_n)
\end{eqnarray}

Where $\mathcal{M}_{pk}$ are the elements of the $\mathcal{M}$ modulation 
matrix with $p=1,...,P$. $\mathcal{M}$ is therefore a $P \times 4$ matrix. 
Before shifting to the next tuning position ($i+1$), HMI will 
measure the next linear combination of Stokes profiles, as given 
by the modulation matrix, at the same wavelength:

\begin{eqnarray}
\mathcal{O}_{p+1}(\lambda_i, t_{n+1}) & = & \sum_{k=1}^{4} 
\mathcal{M}_{p+1,k} \widetilde{I}\obs_k 
(\lambda_i,t_{n+1})
\end{eqnarray}

Once this is done the wavelength position is shifted to $i+1$ and 
the process for $p,p+1$ is repeated until $i=M$. The observation 
sequence continues by changing Equations (1)\,--\,(2) to $p+2,p+3$ 
and so forth until the last scan is made with $P-1,P$. 
Throughout this process, the time index has run from $n=1,...,M \times P$.\\

$\widetilde{\ve{I}}\obs(\lambda_i,t_n)$ results from applying the
HMI filter function at the tuning position $\lambda_i$, 
$\mathcal{G}_i(\lambda_j)$ to the original ({\it i.e.}: solar) Stokes vector 
$\ve{I}\obs(\lambda_j,t_n)$ sampled in a continuous wavelength grid $\lambda_j$ 
at a time $t_n$. 

\begin{eqnarray}
\widetilde{\ve{I}}\obs(\lambda_i,t_n) & = & \int \ve{I}\obs(\lambda_j,t_n)
\mathcal{G}_i(\lambda_j-\lambda_i) {\rm d}\lambda_j
\end{eqnarray}

\begin{figure}
\begin{center}
\includegraphics[width=10cm]{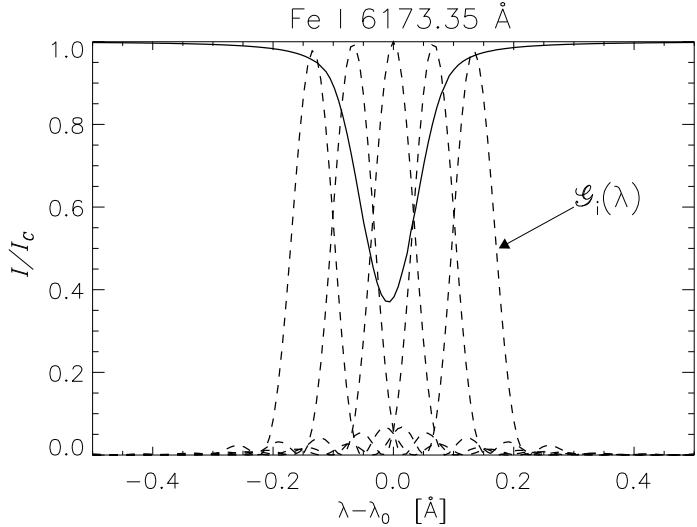}
\caption{HMI filter functions $\mathcal{G}_i(\lambda_j)$ at five
different tuning positions. An example of the intensity
profile for the line Fe {\sc I} 6173.3 \AA~is given in solid.}
\end{center}
\end{figure}

In order to obtain the observed
Stokes vector at each wavelength position, one must first demodulate 
the vector $\ve{O}(\lambda_i)$ from Equation (1). 

\begin{eqnarray}
\widetilde{\ve{I}}\obs(\lambda_i) & = & \mathcal{M}^{-1} 
\ve{O}(\lambda_i)
\end{eqnarray}

or equivalently, as follows.

\begin{eqnarray}
\widetilde{I}_k\obs(\lambda_i) & = & \sum_{p=1}^{P} 
\mathcal{M}_{kp}^{-1} \mathcal{O}_p(\lambda_i,t_n)
\end{eqnarray}

Note that both the vector 
$\ve{O}(\lambda_i)$ and the observed Stokes vector 
$\widetilde{\ve{I}}\obs(\lambda_i)$ (or any of its $k$ components) 
are gathered over 4$PM$ seconds since each element 
$\mathcal{O}_p$ was acquired at a different time $t_n$. Therefore, 
what is observed by HMI is some sort of temporal average of the solar 
Stokes vector.

A number of possible filter sequences and polarization modulations
have been suggested. In this work we will restrict ourselves to
consider only three of them\footnote{All these modulation schemes
take into account that the Doppler camera will scan 
$I\pm V$ in wavelength with a cadence of 50 seconds or better. 
For the vector camera it would have been more desirable
to measure all of polarization states at each tuning position ($\lambda_i$) before 
shifting to the next wavelength. However, this would require too many
motor rotations between observations which may risk mechanical failure.}.   
They will be referred to as: modulations 
A, B or C. Sample framelists for them are given in Tables~I, II, and III.
Referring to those Tables and Equation (1) it is possible to construct the
modulation matrices $\mathcal{M}$ corresponding to modulation schemes A, B, and C.

\begin{equation}
\mathcal{M}_A = \left( \begin{array}{rrrr}
1.000 & 0.810 & 0.000 &0.588 \\
1.000 & -0.810 & 0.000 &0.588 \\
1.000 & 0.000 & 0.810 &-0.588 \\
1.000 & 0.000 & -0.810 &-0.588 \end{array} \right)
\end{equation}

\begin{equation}
\mathcal{M}_B = \left( \begin{array}{rrrr}
1.000 & 0.810 & 0.000 &0.588 \\
1.000 & 0.000 & 0.810 &-0.588 \\
1.000 & -0.810 & 0.000 &0.588 \\
1.000 & 0.000 & -0.810 &-0.588 \end{array} \right)
\end{equation}

\begin{equation}
\mathcal{M}_C = \left( \begin{array}{rrrr}
1 & 1 & 0 &0 \\
1 & -1 & 0 &0 \\
1 & 0 & 1 &0 \\
1 & 0 & -1 &0 \\
1 & 0 & 0 &1 \\
1 & 0 & 0 &-1 \end{array} \right)
\end{equation}

It is important to mention that modulations A and B require 80 seconds to
record the full Stokes vector at all tuning positions, while modulation C
needs 120 seconds. A and B demodulate according to Equations (4)\,--\,(5) since $\mathcal{M}$ 
is a 4$\times$4 square matrix (see Equations (6)\,--\,(7)) and its inverse is calculated easily. 
On the contrary, modulation scheme C requires some other consideration since
$\mathcal{M}_{C}$ (Equation (8)) is not square. This is due to the fact that modulation
C gathers more information than actually needed. In particular, Stokes $I(\lambda)$
is obtained three times. For simplicity, the average $I(\lambda)$ will be used.

\begin{table}
\begin{center}
\caption{Framelist sample for modulation scheme A. The linear
combination of the components of the Stokes vector measured at each time
are given by: $\mathcal{O}_1 = I+aQ+bV$, $\mathcal{O}_2=I-aQ+bV$, 
$\mathcal{O}_3=I+aU-bV$, $\mathcal{O}_4= I-aU-bV$.
Constant values for the linear combinations are: $a=0.81$ and $b=0.588$.
Time is in units of $\Delta t=$ four seconds. $\Delta t$ is determined by CCD readout time, 
the time needed to select a polarization state and to shift the filter position. Note that
the total time needed for modulation A or B to measure the full Stokes vector
across five wavelength points is $PM\Delta t=20\Delta t = 80$ seconds.}
\begin{tabular}{|c|c|c|c|c|c|c|c|c|c|c|c|c|}
\hline
Time/$\Delta t$ [s] & 0 & 1 & 2 & 3 & ... & ... & 2$M$-1 
& 2$M$  \\
\hline
$\mathcal{G}_i$ & 1 & 1 & 2 & 2 & ... & ... & $M$ & $M$ \\
\hline
$\mathcal{O}_p$ & 1 & 2 & 1 & 2 & ... & ... & 1 & 2 \\
\hline
\end{tabular}
\begin{tabular}{|c|c|c|c|c|c|c|}
\hline
2$M$+1 & 2$M$+2 & 2$M$+3 & 2$M$+4 & ... &
4$M$-1 & 4$M$ \\
\hline
 1 & 1 & 2 & 2 & ... & $M$ & $M$ \\
\hline
 3 & 4 & 3 & 4 & ... & 3 & 4 \\
\hline
\end{tabular}
\end{center}
\end{table}

\begin{table}
\begin{center}
\caption{Same as Table~I but for modulation scheme B. }
\begin{tabular}{|c|c|c|c|c|c|c|c|c|c|c|c|c|}
\hline
Time/$\Delta t$ [s] & 0 & 1 & 2 & 3 & ... & ... & 2$M$-1 
& 2$M$  \\
\hline
$\mathcal{G}_i$ & 1 & 1 & 2 & 2 & ... & ... & $M$ & $M$ \\
\hline
$\mathcal{O}_p$ & 1 & 3 & 1 & 3 & ... & ... & 1 & 3 \\
\hline
\end{tabular}
\begin{tabular}{|c|c|c|c|c|c|c|}
\hline
2$M$+1 & 2$M$+2 & 2$M$+3 & 2$M$+4 & ... &
4$M$-1 & 4$M$ \\
\hline
 1 & 1 & 2 & 2 & ... & $M$ & $M$ \\
\hline
 2 & 4 & 2 & 4 & ... & 2 & 4 \\
\hline
\end{tabular}
\end{center}
\end{table}

\begin{table}
\begin{center}
\caption{Framelist sample for modulation scheme C. The linear
combination of the components of the Stokes vector measured each time
are given by: $\mathcal{O}_1 = I+Q$, $\mathcal{O}_2= I-Q$, $\mathcal{O}_3=I+U$, 
$\mathcal{O}_4=I-U$, $\mathcal{O}_5=I+V$, $\mathcal{O}_6= I-V$. Note that
the total time needed by modulation C to measure the full Stokes vector
across five wavelength points is $PM\Delta t=30\Delta t = 120$ seconds.}
\begin{tabular}{|c|c|c|c|c|c|c|c|c|c|c|c|c|}
\hline
Time/$\Delta t$ [s] & 0 & 1 & 2 & 3 & ... & ... & 2$M$-1 
& 2$M$  \\
\hline
$\mathcal{G}_i$ & 1 & 1 & 2 & 2 & ... & ... & $M$ & $M$ \\
\hline
$\mathcal{O}_p$ & 1 & 2 & 1 & 2 & ... & ... & 1 & 2 \\
\hline
\end{tabular}
\begin{tabular}{|c|c|c|c|c|c|c|}
\hline
2$M$+1 & 2$M$+2 & 2$M$+3 & 2$M$+4 & ... &
4$M$-1 & 4$M$ \\
\hline
 1 & 1 & 2 & 2 & ... & $M$ & $M$ \\
\hline
 3 & 4 & 3 & 4 & ... & 3 & 4 \\
\hline
\end{tabular}
\begin{tabular}{|c|c|c|c|c|c|c|}
\hline
4$M$+1 & 4$M$+2 & 4$M$+3 & 4$M$+4 & ... &
6$M$-1 & 6$M$ \\
\hline
 1 & 1 & 2 & 2 & ... & $M$ & $M$ \\
\hline
 5 & 6 & 5 & 6 & ... & 5 & 6 \\
\hline
\end{tabular}
\end{center}
\end{table}

It is also important to note that all schemes have a polarimetric 
efficiency of $\epsilon = 1/\sqrt{3}$ when measuring Stokes $Q$, $U$, and $V$.
According to del Toro Iniesta and Collados (2000) this translates into an error
(due to photon noise) $\sigma_i$:

\begin{equation}
\sigma_i^2 = \frac{1}{P}\frac{\sigma^2}{\epsilon^2}
\end{equation}

\noindent where $\sigma^2$ is roughly the noise level in $\mathcal{O}_p$ (defined in Equation (1)). 
Note that scheme C, where $P=6$, introduces a factor $2/3$ less noise than schemes 
A and B ($P=4$). However, scheme C takes a factor $3/2$ longer to be completed 
(120 seconds instead of 80 seconds) so that the noise in the Stokes parameters $\sigma_i$ per unit time
is the same for all three schemes.

\section{Simulated HMI Observed Profiles}%

In order to simulate Stokes profiles as if they were observed by the HMI
vector camera, a set of parameters [$\ve{X}_{\rm r}(t_n)$] describing the conditions present
in the solar photosphere at a time $t_n$, is chosen. In the following we will assume that
the solar photosphere is described by a Milne-Eddington (ME) atmosphere
({\it i.e.}: physical parameters are constant with depth in the photosphere; see
Landi Degl'Innocenti (1991) for details). This set of parameters is
presented in Table IV. $\ve{X}_{\rm r}(t_n)$ is used to solve the 
radiative transfer equation (RTE) according to the ME analytical
solution given by Landi Degl'Innocenti (1991) or Del Toro Iniesta (2003).
This provides the Stokes vector emerging from a solar magnetized atmosphere
 [$\ve{I}\mg(\ve{X}_{\rm r})$]. In addition, a contribution from a non-magnetic
atmosphere ($\ve{I}\nm$) is added using the filling factor $\alpha$ 
(fractional area of the resolution element
occupied by the magnetic atmosphere) in order to produce the simulated
observed profiles:

\begin{equation}
\ve{I}\obs (\lambda_j,t_n) = \alpha \ve{I}\mg(\lambda_j,\ve{X}_{\rm r}[t_n]) 
+ (1-\alpha) \ve{I}\nm(\lambda_j)
\end{equation}

\noindent or equivalently, taking into account that the non-magnetic atmosphere
does not produce polarization signals:

\begin{equation}
\left(\begin{array}{l}
I\obs \\ Q\obs \\ U\obs \\ V\obs \end{array} \right) =
\alpha
\left(\begin{array}{l}
I\mg \\ Q\mg \\ U\mg \\ V\mg \end{array} \right)
+ (1 - \alpha)
\left(\begin{array}{l}
I\nm \\ 0 \\ 0 \\ 0 \end{array} \right)
\end{equation}

\begin{table}
\caption[]{Set of Milne-Eddington physical parameters $\ve{X}_{\rm r}$ used
to produce synthetic profiles.}
\begin{center}
\begin{tabular}{|l|c|c|}
 \hline
Physical Parameter & Units & Symbol\\
\hline
\hline
Magnetic field strength & Gauss & B\\
Magnetic field inclination & deg & $\gamma$\\
Magnetic field azimuth & deg & $\psi$\\
Line-of-sight velocity & km s$^{-1}$ & v$^{\rm LOS}$\\
Source Function & dimensionless & S\\
Source Function Gradient & dimensionless & $\rm \dot{S}$\\ 
Filling Factor & dimensionless & $\alpha$\\
\hline
\end{tabular}
\end{center}
\end{table}

$\ve{I}\obs(\lambda_j,t_n)$ is sampled using a
grid of 300 points in wavelength (as an approximation
to a continuous function; see Equation (3)). This Stokes vector enters into
Equation (3), where the HMI filter profiles $\mathcal{G}_i$ are applied
and $\tilde{\ve{I}}\obs(\lambda_i,t_n)$ (sampled with a coarse grid of five
wavelength points) is obtained. According to the selected polarization modulation,
 $\mathcal{M}$, a linear combination of the components of the Stokes vector, 
 $\mathcal{O}_p(\lambda_i,t_n)$, is produced (see Equation (1)).

\subsection{PHOTON NOISE}%
Considering that the full-well depth of the CCD camera is approximately
100 000 detected photons, it is reasonable to limit the exposure time in order to
reach 4/5 of that value in order to avoid saturation effects.
Photon noise $\sigma$ is added into $\mathcal{O}_p(\lambda_i,t_n)$ considering
that 4/5 of the CCD full well depth is achieved for the brightest
solar feature ({\it i.e.}: quiet Sun at disk center). For profiles intended to
simulate other solar features, {\it e.g.} sunspots, photon noise is added
according to the expected intensity level. The noise is added
using random number generation following Poisson's statistics.
This is often referred to as {\it shot noise}. 

Note that this noise level $\sigma$ translates into different noise for 
the polarization signals $\sigma_i$ (see Equation (9)). Since we add noise into
$\mathcal{O}_p(\lambda_i)$, $\sigma_i$ propagates naturally after demodulating 
(Equations (4)\,--\,(5)).

\subsection{SOLAR  P MODES}%
The readout time of the 4096 $\times$ 4096 CCD camera is about 2.7 seconds. 
Together with the time required to rotate the waveplates and some idle time, the total time 
lapse before $\mathcal{O}_{p+1}(\lambda_i,t_n+\Delta t)$ is measured is: $\Delta t=$ four seconds. 
In this time interval the physical conditions in the solar photosphere might have changed:
$\ve{X}_{\rm r}(t_n+\Delta t) \ne \ve{X}_{\rm r}(t_n)$, and therefore
the observed $\ve{I}\obs(\lambda_j,t_n+\Delta t)$ is likely to be different
from $\ve{I}\obs(\lambda_j,t_n)$.

There are many processes responsible for the
variation of the physical conditions of the solar photosphere
on time scales of seconds. In this paper we will focus of the effects
of the {\it p} modes only. This means that from the set of parameters
$\ve{X}_{\rm r}(t_n)=[\rm{B},\gamma,\psi,v^{\rm LOS}, {\rm S}, \dot{\rm S}]$,
only $\rm v^{LOS}$ will be considered to change in time. In particular:

\begin{equation}
\rm v^{\rm LOS}(t) = \rm v^{\rm LOS}_{\rm plasma}+\xi(t)
\end{equation}

\noindent where $\rm v^{\rm LOS}_{\rm plasma}$ corresponds to a real plasma velocity,
while $\xi(t)$ is the velocity associated with the solar {\it p} modes. An example
of the function $\xi(t)$ is presented in Figure 2 (top panel). This was obtained
from a two hour average of quiet-Sun MDI data. The {\it rms} amplitude of the oscillation
is about 250 m s$^{-1}$.

\begin{figure}
\begin{center}
\includegraphics[width=12cm]{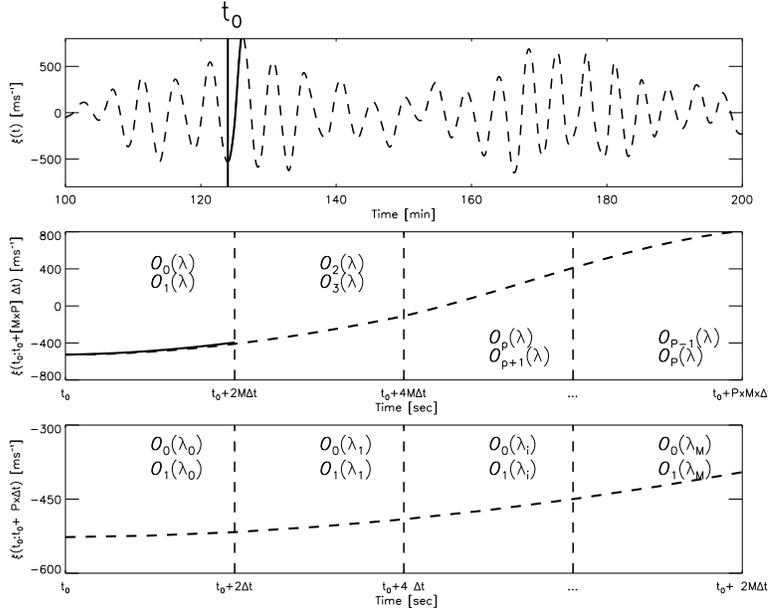}
\caption{{\it Top panel}: dashed line describes the function $\xi(t)$ (Equation (12))
representative of the solar {\it p} modes as measured by the MDI high resolution field 
of view. The vertical solid line indicates a randomly chosen time $t_0$ at which 
HMI starts acquiring data. The solid line indicates the variation of the solar
{\it p}-mode velocity during the time interval $[t_0,t_0+PM\Delta t]$
. This is the time required to record the full Stokes vector across $M$ tuning positions.
{\it Middle panel:} close up of the function $\xi(t)$ over the data acquisition time 
interval (solid line in the upper panel). The vertical dashed lines separate
the time intervals in which $\mathcal{O}_p(\lambda)$ and 
$\mathcal{O}_{p+1}(\lambda)$ are recorded at all tuning positions,
taking a total of $[t_0,t_0+ 2M\Delta t]$ 
indicated by the solid line. 
{\it Bottom panel}: close up of the solid line in the middle panel. 
Each vertical dashed line separates the time interval in which 
$\mathcal{O}_p$ and $\mathcal{O}_{p+1}$ are taken at each tuning position
($\lambda_i$). Between $\mathcal{O}_p$ and $\mathcal{O}_{p+1}$, there is a 
time delay $\Delta t$.}
\end{center}
\end{figure}

\paragraph{Long time-scale p-mode effect}
A close inspection of Figure 2 (middle panel) indicates that, for example
using modulation scheme C, during the time interval $[t_0,t_0+2 M \Delta t]$
the HMI vector camera will measure: $\mathcal{O}_0(\lambda)=I(\lambda)+Q(\lambda)$
and $\mathcal{O}_1(\lambda)=I(\lambda)-Q(\lambda)$. This allows an acquisition of both
$I(\lambda)$ and $Q(\lambda)$. During the next time interval 
$[t_0+2 M \Delta t,t_0+4 M \Delta t]$ HMI records: 
$\mathcal{O}_2(\lambda)=I(\lambda)+U(\lambda)$ and $\mathcal{O}_3(\lambda)
=I(\lambda)-U(\lambda)$. This provides $I(\lambda)$ and $U(\lambda)$. Due to
the temporal evolution of the solar {\it p}-mode velocity [$\xi(t)$] during this time,
$Q(\lambda)$ and $U(\lambda)$ are affected, on average, by different
velocities and therefore they will be shifted in wavelength with respect to each other.
Hereafter this effect will be referred to as the long time scale {\it p}-mode effect.

\paragraph{Short time scale p-mode effect}
Another effect associated with the solar {\it p} modes can be visualized by
looking at Figure 2 (bottom panel). The time $t_0$ is when the vector camera 
records $\mathcal{O}_p(\lambda_i)$. Immediately after:
$t_0+\htt$, is when HMI measures the state $\mathcal{O}_{p+1}(\lambda_i)$:
Let us consider for example that modulation scheme C is used. In this case
$\htt = \Delta t =$ four seconds. We can thus write:

\begin{eqnarray}
\mathcal{O}_{2p-1}(\lambda_i,t_0) & = & I\obs_1(\lambda_i,t_0) + I\obs_{p+1}(\lambda_i,t_0)
 \;\;\;\;\;\;\;\;\; m = 1,2,3
\end{eqnarray}

and a time $\htt$ later the orthogonal state is measured:

\begin{eqnarray}
\mathcal{O}_{2p}(\lambda_i,t_0+\htt) & = & I\obs_1(\lambda_i,t_0+\htt) 
- I\obs_{p+1}(\lambda_i,t_0+\htt)
\end{eqnarray}

In order to isolate the polarization signal $I\obs_m$ one would normally proceed as
follows:

\begin{eqnarray}
\begin{split}
\mathcal{O}_{2p-1}(\lambda_i,t_0)-\mathcal{O}_{2p}(\lambda_i
,t_0+\htt) & =  I\obs_1(\lambda_i,t_0)-I\obs_1
(\lambda_i,t_0+\htt)\\
& +I\obs_{p+1}(\lambda_i,t_0)+I\obs_{p+1}(\lambda_i,t_0+\htt)
\end{split}
\end{eqnarray}

Both $I_1$ and $I_m$ have changed between $t_0$ and $t_0+\htt$.
The intensity profiles ({\it i.e.}: Stokes $I$ or $I_1$) have a much larger
amplitude that the polarization profiles $I_m$. Therefore we can ignore changes 
in $I_m$ and consider that the solar {\it p} modes have shifted only the intensity profiles:

\begin{eqnarray}
\mathcal{O}_{2p-1}(\lambda_i,t_0)-\mathcal{O}_{2p}(\lambda_i
,t_0+\htt) & \simeq  2 I\obs_{p+1}(\lambda_i,t_0)+\underbrace{\left[\frac{\partial 
I\obs_1(\lambda_i,t)}{\partial t}\right]_{t=t_0} \htt}_{{\rm xtalk} \;\;\; I_1 \rightarrow I_m} 
\end{eqnarray}

\noindent where the term indicated as {\it xtalk} in Equation(16) will be referred to as short time-scale
{\it p}-mode effect. An order-of-magnitude estimation of this effect can be
made by modeling the intensity profile as a Gaussian and the temporal
evolution of the solar {\it p} modes in the following form:

\begin{eqnarray}
I_1\obs(\lambda,t) & = & 1-I_C \exp{\left(-\beta[\lambda
-\lambda_0(1-\frac{\xi(t)}{c})]^2\right)}\\
\xi(t) & = & V_0 \cos(\frac{2\pi t}{T})
\end{eqnarray}

We can easily calculate the cross talk from Stokes $I$ into
the polarization profiles as:

\begin{eqnarray}
\frac{\partial I\obs_1(\lambda_i,t)}{\partial t} & = &
\frac{2\lambda_0\beta}{c}[1-I\obs_1(\lambda,t)]\left[\lambda-
\lambda_0(1-\frac{\xi(t)}{c})\right]\frac{\partial \xi(t)}{\partial t}
\end{eqnarray}

Using the following typical values for the neutral iron line Fe {\sc I}
6173.3 \AA~taken from Graham {\it et al.} (2002, 2003) and Norton {\it et al.} (2006):
$I_C = 0.5$, $V_0 = 250 $ m s$^{-1}$ (see Figure 2; top panel),
$\beta = 385 $ \AA$^{-2}$, and $T = 300$ s, we have evaluated
Equation (19) and plotted it in Figure 3 in units of percentage of
the continuum intensity. As expected, the maximum cross talk 
is produced in the line wings because that is where changes in $I_1$ 
are maximum when the velocity changes. The maximum cross talk is 
produced when the {\it p}-mode velocity changes sign. Figure 3 also
shows that along the time axis the cross talk changes sign for
a particular wavelength ({\it e.g.}: in the increasing region
of the {\it p}-mode velocity the cross talk has opposite sign than
in the decreasing velocity region), implying that integrating over time
can reduce the cross talk level (see Section 6).
The gray scale in Figure 3 is set so that, in the best case scenario, where 
$\htt = \Delta t =$ four seconds (modulation scheme C), we have an estimated cross 
talk of $\sim$ 0.15 $\times$ 4 $=$ 0.6 \%, while in other cases, where $\htt \sim 50-80$ 
seconds, the amount of cross talk could be as much as 7\,--\,9 \%.   The modulation matrix 
$\mathcal{M}$ in Equations (6)\,--\,(8) defines how far away in 
time the measurements $\mathcal{O}_p$ (that will be combined
to produce each Stokes profile) are carried out and therefore defines $\htt$.\\

\begin{figure}
\begin{center}
\includegraphics[angle=0,width=10.5cm]{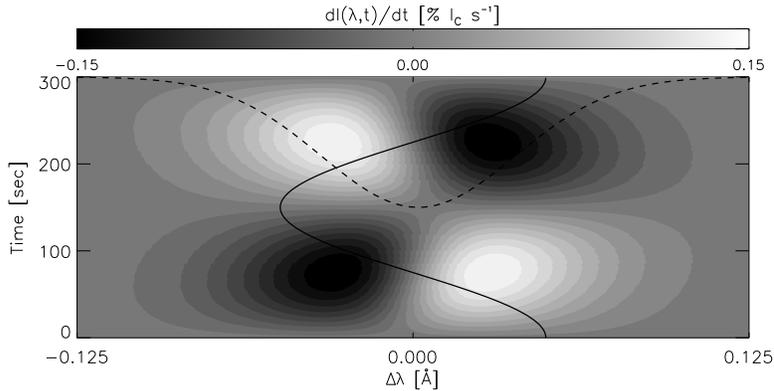}
\end{center}
\caption{Contour plot of the amount of cross talk 
per unit time (in units of \% of the continuum intensity)
induced into the polarization profiles from the
total intensity as a consequence of the solar
{\it p} modes (see Equation (19) and text for details). Horizontal axis
indicates relative position with respect to the central
laboratory wavelength. Vertical axis indicates time
in which the measurement is taken within one oscillatory 
period of 300 seconds. Relative amplitudes are indicated by the
gray scale bar at the top of the plot. For visualization purposes,
a magnified example of the Fe {\sc I} 6173 \AA~ spectral line has
been over plotted in dashed lines. Temporal evolution of
the {\it p}-mode velocity $\xi(t)$ (see Equation (18)) is indicated by
the solid lines.}
\end{figure}

Besides this analytical estimate, we have carried out a more detailed
analysis by taking into account all possible effects and a realistic
{\it p}-mode function $\xi(t)$ (Figure 2 top panel). For this analysis we
used 20 000 Stokes profiles $\widetilde{\ve{I}}\obs(\lambda_j,t_n)$
that were converted into $\widetilde{\ve{I}}\obs(\lambda_i)$
according to Section 2. Polarization levels, defined as
${\rm max}|V|$ and ${\rm max}|\sqrt{Q^2+U^2}|$ in units
of percentage of the continuum intensity, are compared before and after.
Results are displayed in Figure 4. The horizontal axis refers
to the polarization levels from the original profiles: 
$\widetilde{\ve{I}}\obs(\lambda_i,t_n)$ as in Equation (3). The vertical axis
represents the polarization levels of the profiles once the effects 
of the solar {\it p} modes have been taken into account: $\widetilde{\ve{I}}\obs(\lambda_i)$
as in Equation (4). Random initial times $t_0$ are used for sampling the function $\xi(t)$ (see Figure 2; top
panel). The experiment is repeated for all modulation schemes (see Section 2).
As can be seen, modulation scheme A introduces a large amount of cross talk into the circular
polarization while keeping the cross talk into the linear polarization within reasonable limits.

As already mentioned, this effect is ultimately related to the modulation matrix $\mathcal{M}$.
In scheme A, $\mathcal{O}_1(\lambda_i,t_0)=I - 0.81 Q + 0.588 V$ it is the first
measurement. The next one corresponds to $\mathcal{O}_2(\lambda_i,t_0+\Delta t)=I 
- 0.81 Q + 0.588 V$ and therefore,
Stokes $Q(\lambda_i)$ can be constructed with measurements done only $\Delta t =$ four seconds 
apart from each other, thereby minimizing the amount of cross talk from Stokes $I$. 
However, Stokes $V$ is built out of a combination of measurements taken far away in time. For example:
$\mathcal{O}_1(\lambda_i,t_0)=I + 0.81 Q + 0.588 V$ is measured at $t_0$, but 
$\mathcal{O}_3(\lambda_i,t_0+[2M+1]\Delta t)=I + 0.81 U - 0.588 V$ is recorded 
44 seconds later (see Table~I). During this time the solar {\it p} modes might introduce large levels
of cross talk from Stokes $I$ into Stokes $V$. This also explains 
why scheme B introduces large cross talk into the linear polarization while
leaving the circular polarization almost unchanged. As expected, Modulation scheme 
C performs well both in linear and circular polarization. The obvious drawback to
Scheme C is that it takes longer to scan across the line: $6M\Delta t$ seconds as compared 
to $4M\Delta t$ seconds from schemes A and B. Thus, scheme C minimizes the short 
time scale {\it p}-mode effect but maximizes the long time scale {\it p}-mode effect. In order 
to study which effect introduces larger errors when retrieving the magnetic field vector, 
inversion of simulated profiles are required.

\begin{figure}
\begin{center}
\includegraphics[angle=0,width=10.5cm]{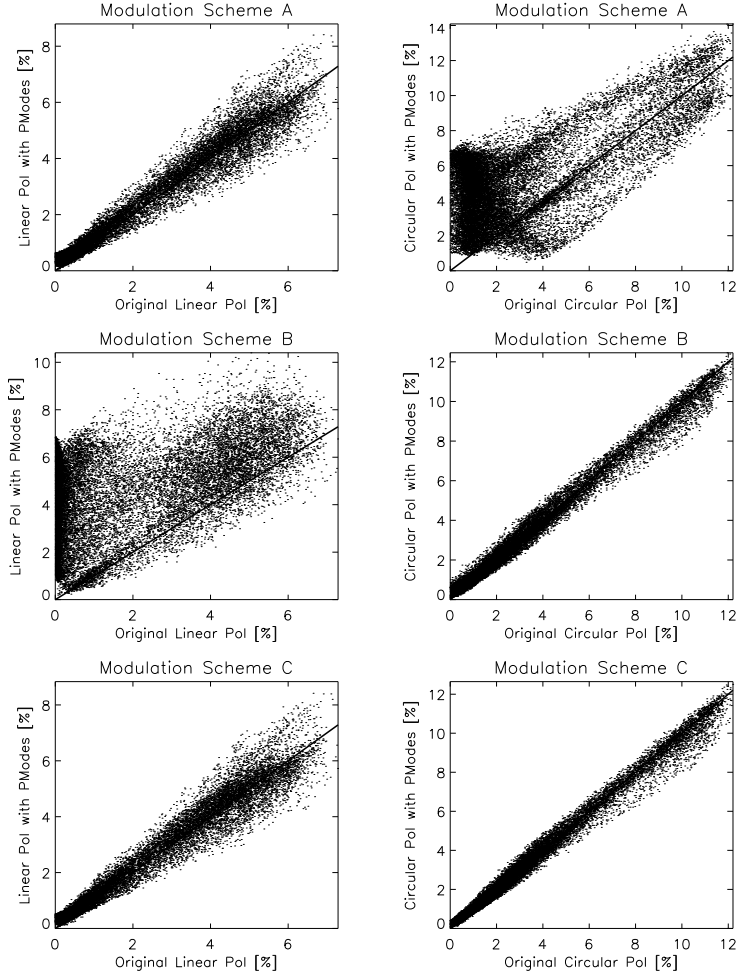}
\end{center}
\caption{Polarization cross talk induced by the short time solar {\it p}-mode effect for
the three polarization modulations under consideration: scheme A (top panels), scheme
B (middle panels) and scheme C (bottom panels). Horizontal axis refer to the original
polarization levels (in percentage of the continuum intensity units). Vertical axis
indicate the polarization levels after the {\it p}-mode effect has been taken into account.
Left panels are for the linear polarization: $\sqrt{Q^2+U^2}$. Right panels
for the circular polarization: $V$. See text for further details.}
\end{figure}

\section{Inversion strategy}%

\subsection{OBSERVED PROFILES}%
Prior to the inversion, we must obtain a sufficiently large number
of observed Stokes profiles whose atmospheric parameters are known.
To this end we have built up a data base of atmospheric parameters
with more than 20 000 atmospheres. Each of those is represented by
a set of parameters $\ve{X}_{\rm r}=[{\rm B},\gamma,\psi
,{\rm v}^{\rm LOS}, {\rm S}, {\rm \dot{S}}, \alpha]$. Each $\ve{X}_{\rm r}$
is used to produced a synthetic (that simulates observed ones) Stokes vector: 
$\ve{I}\obs(\lambda_j)$ sampled in 300 wavelength positions to
approximate a continuos function (Equation (3)). This is done according to
Equation (10). Note that $\ve{X}_{\rm r}$ is used to synthesize the magnetic atmosphere:
$\ve{I}\mg(\lambda_j)$. To this profile we add a non-magnetic contribution 
$\ve{I}\nm(\lambda_j)$ whose shape is known.

A random time ($t_0$) and a modulation matrix ($\mathcal{M}$) must also be chosen.
According to this and Equation (3) the HMI filter profiles $\mathcal{G}_i$ are applied
and $\widetilde{\ve{I}}\obs(\lambda_i,t_0)$ is obtained at wavelength position $\lambda_i$.
A linear combination of the components of $\widetilde{\ve{I}}\obs(\lambda_i,t_0)$
is produced through Equation (1): $\mathcal{O}_p(\lambda_i,t_0)$. The next
linear combination $\mathcal{O}_{p+1}(\lambda_i,t_0+\Delta t)$ at the same wavelength position
is obtained out of $\widetilde{\ve{I}}\obs(\lambda_i,t_0+\Delta t)$, where these
profiles have been shifted in wavelength to account for the variation over that time
interval $\Delta t$ of the function $\xi(t)$ (see Equation (12) and Figure 2). Note that
the only velocity that changes in time is the {\it p}-mode velocity $\xi(t)$. In Equation (12)
${\rm v}^{\rm LOS}_{\rm plasma}$ corresponds to ${\rm v}^{\rm LOS}$ from $\ve{X}_{\rm r}$
and does not change during the observing time. Unless otherwise specified, photon noise 
is also added to each $\mathcal{O}_p(\lambda_i,t_0)$ according to Section 3.1.

In Figure 2, the {\it rms} amplitude of the {\it p}-mode velocity is about 250 m s$^{-1}$. This was
obtained from a two-hour MDI temporal series in the quiet Sun using the high resolution field
of view. For magnetized regions the amplitude of the solar {\it p} modes is known to decrease, however. To this end, the
rms value of the $\xi(t)$ function in Figure 2 is scaled with the magnetic field $\rm B$
from $\ve{X}_{\rm r}$ according to (see Norton, 2000):

\begin{eqnarray}
\xi_{rms}(t) & = & \left\{\begin{array}{rr}
250 - 0.155 \times {\rm B} & \;\;\;\; {\rm if} \;\;\; {\rm B} \le 1000~G \\
128 - 0.033 \times {\rm B} & \;\;\;\; {\rm if} \;\;\; {\rm B} > 1000~G  \\
\end{array}\right\} \;\;\; {\rm [m\;s^{-1}]}
\end{eqnarray}

Once all needed $\mathcal{O}_p(\lambda_i)$, with $p=1,...,P$, are known at wavelength 
position $\lambda_i$, the final observed vector $\widetilde{\ve{I}}\obs(\lambda_i)$
is obtaining by demodulating as in Equations (4) - (5). An example of $\ve{I}\obs(\lambda_j,t_0)$ is given in
Figure 5 (top four panels; dashed line). In circles we also plot the observed profiles,
$\widetilde{\ve{I}}\obs(\lambda_i,t_0)$, after the filters $\mathcal{G}$ have been applied. 
In the lower four panels in Figure 5, $\widetilde{\ve{I}}\obs(\lambda_i,t_0)$ is plotted again. Here
we added $\widetilde{\ve{I}}\obs(\lambda_i)$, which are the
observed profiles after the effects of the solar {\it p} modes have been introduced. Modulation 
schemes A (dotted lines), B (dashed lines) and C (dashed-dotted lines) have been used. 
In the three cases, the same $t_0$ was employed. This plot also shows how the solar {\it p} modes
affect the circular polarization more when scheme A is used, and have a stronger effect in
the linear polarization profiles when scheme B is employed.

\begin{figure}
\begin{center}
\includegraphics[angle=0,width=9cm]{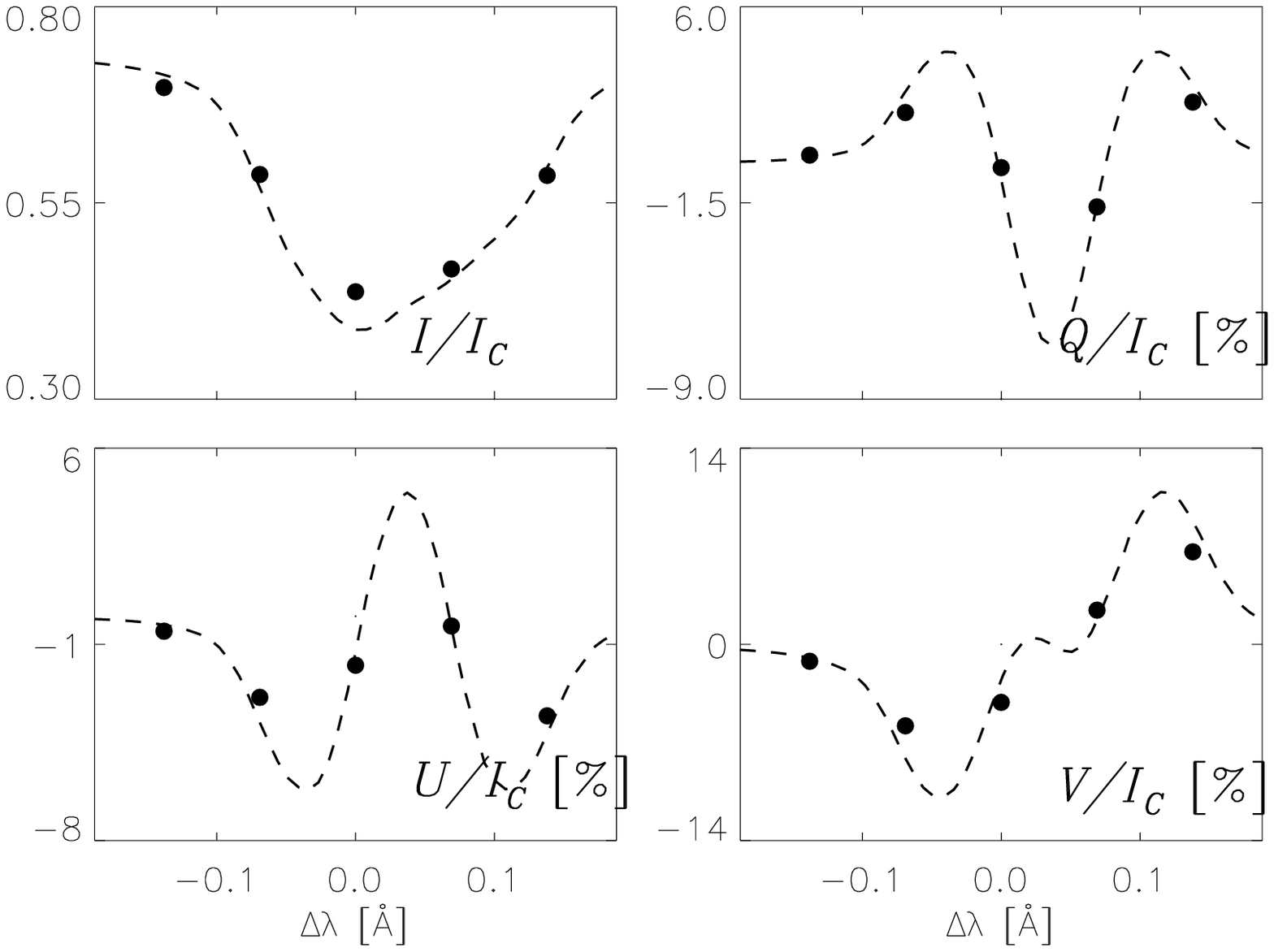} \\
\includegraphics[angle=0,width=9cm]{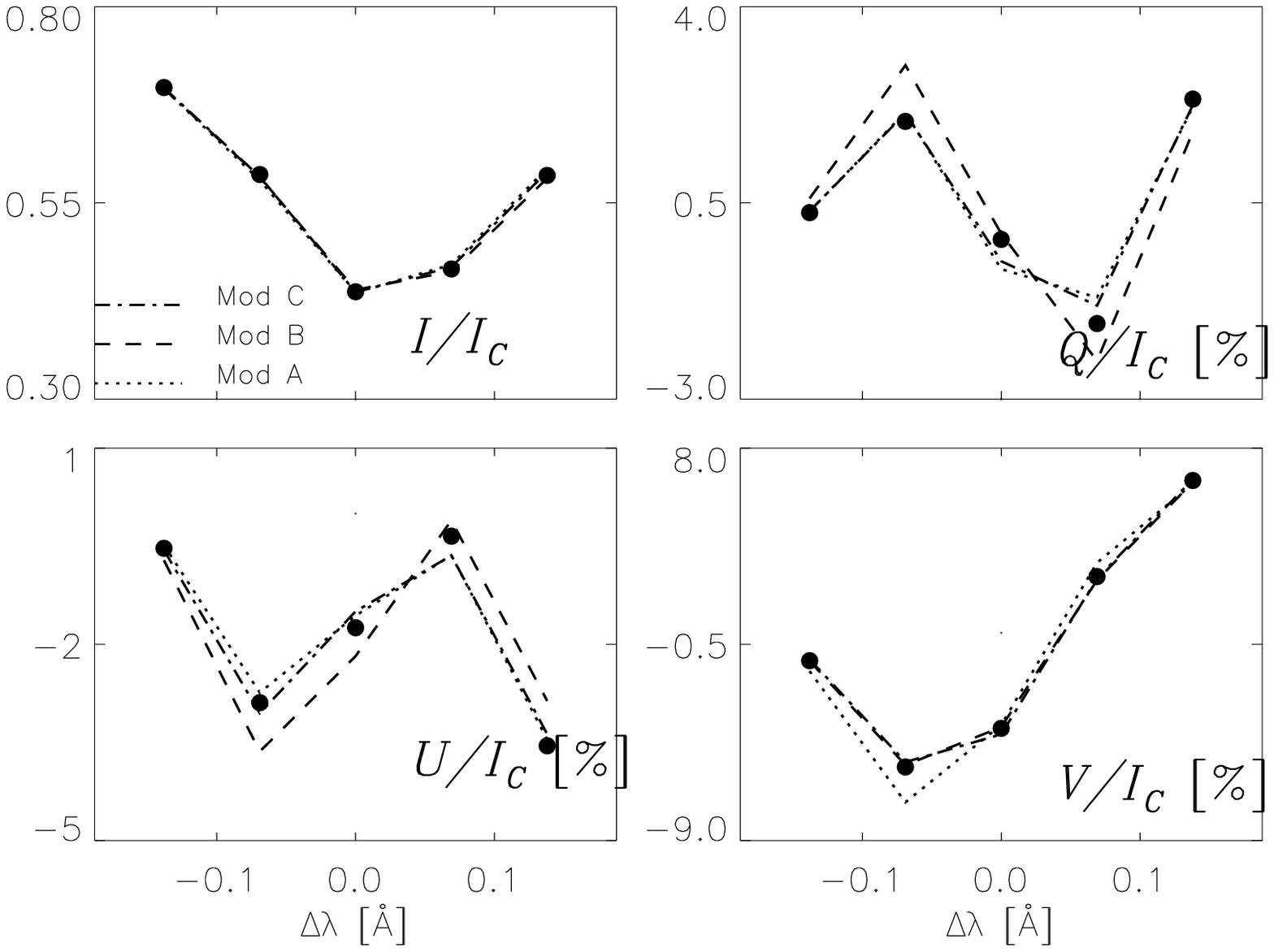}
\end{center}
\caption{{\it Top 4 panels}: Normalized original Stokes profiles $\ve{I}\obs(\lambda_j)$
evaluated in 300 wavelength points (dashed lines). Profiles $\widetilde{\ve{I}}\obs(\lambda_i)$ 
after convolution with HMI filter profiles (Equation (3); circles). {\it Bottom 4 panels}: circles as
in the top panels. Dotted, dashed and dashed-dotted lines corresponds to the profiles observed
by HMI, $\widetilde{\ve{I}}\obs(\lambda_i)$ (see  Equation (4)), when modulation schemes A, B, and C,
respectively, are used. No photon noise has been considered here.}
\end{figure}

\subsection{INVERSION DETAILS}%

Once HMI observed profiles $\widetilde{\ve{I}}\obs(\lambda_i)$ are available,
the next step is to try to recover the photospheric parameters $\ve{X}_{\rm r}$ that produced them.
The underlying idea of this inverse process consists in producing a initial synthetic Stokes
profiles $\widetilde{\ve{I}}\syn(\lambda_i)$ out of an initial set of photospheric
parameters $\ve{X}_{\rm 0}$ by solving the analytical Milne-Eddington solutions to the
RTE. The synthetic profiles are compared to the observed ones and
 $\ve{X}_{\rm 0}$ is modified until the observations are matched.
The way this works is by building a $\chi^2$ or merit function as follows:

\begin{eqnarray}
\chi^2 & = & \frac{1}L\sum_{k=1}^{4} \sum_{i=1}^{M} \left[\frac{\widetilde{I}_k^{\rm
  obs}(\lambda_i)-I_k^{\rm syn}(\ve{X},\lambda_i)}{\sigma_{ik}}\right]^2 w_k^2
\end{eqnarray}

\noindent where $L$ is the total number of free parameters ({\it i.e.}: seven parameters defining
$\ve{X}_{\rm r}$), and the indexes  {\it k} and {\it i} run for the four components of 
the Stokes vector and for all wavelengths in which the spectral line is sampled $M=5$. 
$\sigma_{ik}$ is a measure of the errors in the observations ({\it i.e.}: photon noise) 
and $w_k$ is a factor used to ensure that the contribution on the derivatives of $\chi^2$
with respect to the free parameters is not dominated by derivatives with respect to Stokes $I$.
To this end, we adopt:

\begin{eqnarray}
w_k^2 & = & \frac{1}{max\{[I_k(\lambda)^2]\}}
\end{eqnarray}

Once $\chi^2$ is built, we numerically calculate the derivatives of the 
merit function with respect to the free parameters which are then used by
a non-linear Levenberg-Marquardt least squares fitting algorithm (Press {\it et al}. 1986)
together with a modified Singular Value Decomposition method (Ruiz Cobo and 
Del Toro Iniesta, 1992). The algorithm returns the vector of free parameters 
\ve{X}$_{\rm f}=[\rm B_{f},\gamma_{f},\psi_{f},V^{LOS}_{f},S_{f},\dot{S}_{f},\alpha_{f}]$
that minimizes $\chi^2$ and defines the actual ({\it i.e.}: real) solar atmosphere:
$\ve{X}_{\rm r}=\ve{X}_{f}$.

As pointed out, the inversion requires the use of an initial guess model $\ve{X}_0$. In order
to make the simulation as realistic as possible, we choose
to fix the initial set of parameters (see Table~V) that will be kept
for all inverted profiles. In that way we assume no previous knowledge
of the real $\ve{X}_{\rm r}$.  Besides the initial set of parameters ($\ve{X}_0$) 
we run a number of extra inversions of each Stokes vector but 
using randomly generated $\ve{X}$'s, and only the one that retrieves 
the smallest $\chi^2$ is kept.

It is important to mention that our set of free parameters (see Table V) does not include
the Doppler width, line-to-continuum opacity ratio and damping (parameters also used
by the Milne-Eddington model). This strategy has been adopted to mimic the number 
of free parameters of a full inversion method (see for instance Ruiz Cobo and
Del Toro Iniesta, 1992) where the neglected parameters are determined through the 
temperature stratification $T(\tau)$.

\begin{table}
\caption[]{Initial set of physical parameters \ve{X}$_0$ used in the inversion.
We only assume a previous knowledge in the magnetic field inclination, where
$\gamma < 90^{\circ}$ or $>90^{\circ}$ is chosen depending on the sign in
Stokes $V$ lobes. Note that the Doppler width, opacity ratio and damping parameters 
are not considered as free parameters (see text for details).}
\begin{center}
\begin{tabular}{|l|c|c|}
 \hline
Physical Parameter & Symbol & Value\\
\hline
\hline
Magnetic field strength & B & 1000 Gauss\\
Magnetic field inclination & $\gamma$ & 60 - 120$^{\circ}$\\
Magnetic field azimuth & $\psi$ & 30$^{\circ}$\\
Line-of-sight velocity & v$^{\rm LOS}$ & 2.5 km s$^{-1}$\\
Source Function & S & 0.3 \\
Source Function Gradient& $\rm \dot{S}$ & 0.45\\ 
Filling Factor & $\alpha$ & 0.5\\
\hline
\end{tabular}
\end{center}
\end{table}

\section{Effect of Polarization Modulation}%

In this section we will study which polarization scheme allows for a better recovery
of the physical photospheric parameters $\ve{X}_{\rm r}$ and in particular of the
magnetic field vector. To this end, three simulations (see Section 4 for details) were carried
out using a different modulation matrix $\mathcal{M}$ (see Equations (6)\,--\,(8)). Each of them 
includes the inversion of 20 000 individual profiles. Those profiles were 
produced using a set of photospheric parameters ($\ve{X}_{\rm r}$), and a initial 
random time $t_0$ to take into account the solar {\it p} modes (as in Figure 2; top panel). 
The {\it p}-mode function $\xi(t)$ was scaled according to Equation (20). 
At this point, photon noise was not taken into account. The inversion of each profile
returns a set of inverted atmospheric parameters ($\ve{X}_{\rm f}$) that can be compared
to the real one ($\ve{X}_{\rm r}$). This allows us to study the error in the retrieval
of those parameters. 

Figure 6 displays the errors in the retrieval of magnetic field strength (top left panel), 
magnetic field inclination (top right panel), magnetic field azimuth (bottom right panel) 
and line-of-sight velocity (bottom left panel). The errors are presented as a function of 
the physical parameter itself. This measure is taken by sorting the differences 
between the real and inverted values in increasing order
and picking the value that leaves 68 \% of the 20 000 inverted profiles below it.
This provides a robust measure of the standard deviation in the inferred parameters
and will be used hereafter as error.

In spite of having the longest observing sequence ($30\Delta t$ seconds),
scheme C minimizes the errors in the retrieved parameters.
This indicates that the short term {\it p}-mode effect is more important than
the long one (see Section 3.2). In this sense, scheme C ensures that the 
induced cross talk is minimized since orthogonal observations are done with 
only $\hat{t}=\Delta t=$ four seconds delay (see Equations (13)\,--\,(14)).

\begin{figure}
\begin{center}
\includegraphics[angle=0,width=12cm]{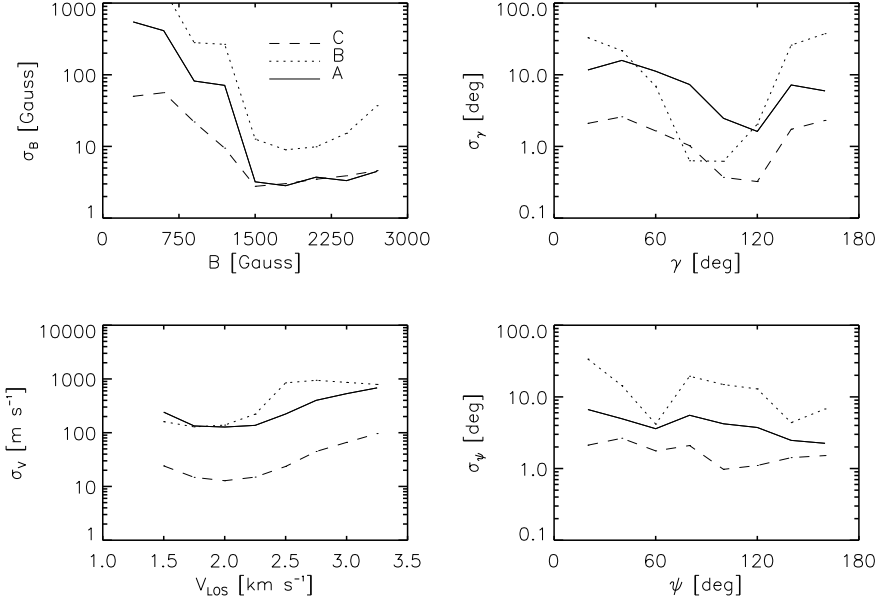}
\end{center}
\caption{Errors in the retrieval of the different physical parameters
as a function of the parameter itself. {\it Top left panel}: magnetic field
strength; {\it Top right panel}: magnetic field inclination; 
{\it Bottom left panel}: line-of-sight velocity; {\it Bottom left panel}: 
magnetic field azimuth. Different line styles correspond to: modulation
scheme A (solid; Equation (6)), scheme B (dotted; Equation (7)), scheme C (dashed; Equation (8)).}
\end{figure}

\begin{itemize}
\item {\it Magnetic field strength}: scheme A does a better job that scheme B. The main
 reason for this is to be found in the generally smaller amplitudes of Stokes $Q$ and $U$ 
with respect to Stokes $V$. The same amount of cross talk will alter 
more significantly the linear polarization than the circular, thereby affecting in a more 
severe manner the recovered values for $B_{\perp}$ in scheme B than the retrieved 
$B_{\parallel}$ in scheme A. Note that all schemes show an increase in the
errors when the field strength grows above 1500 Gauss. This is related to 
{\it saturation effects}: for larger fields, the Zeeman splitting is large enough 
to shift the spectral line out of the scanned region, resulting in a loss of information.
For small field strengths the high increase in the errors, in particular for schemes A and B, 
is mainly due to the larger amplitude of the {\it p} modes (Equation (20)).
\item {\it Magnetic field azimuth}: as expected, scheme A retrieves smaller errors 
than scheme B. The reason is that the field azimuth angle 
depends on the ratio $\sim Q/U$ (Jefferies and Mickey, 1991). 
Since scheme A minimizes the
 {\it p}-mode cross talk into the linear polarization (see Figure 4), it retrieves 
more accurate azimuthal angles. 
\item {\it Magnetic field inclination}: scheme B gives slightly superior 
performance as compared to scheme A. This has no trivial explanation, 
since this parameter depends on both linear and circular polarization.
In addition, all three modulation schemes display this kind of behavior in $\sigma_\gamma$
as a function of $\gamma$. The error increases towards $\gamma \rightarrow 
0^{\circ},180^{\circ}$, while it is minimum at $\gamma \sim 90^{\circ}$.
\end{itemize}

Another interesting comparison can be made by plotting the error
as a function of the filling factor of the magnetic component 
($\alpha$). This is done in Figure 7. Errors increase with decreasing
$\alpha$ for all physical parameters and for all modulation schemes.
The reason in this case is trivial. The closer $\alpha$ is to 0, the
less amount of total observed light comes from the
magnetic atmosphere $\ve{I}\mg$ (see Equations (10)\,--\,(11)), therefore becoming 
increasingly more difficult to infer its properties. Hereafter we will focus
in the modulation schemes A and C, since scheme B can be ruled out due
to the larger errors that introduces.

\begin{figure}
\begin{center}
\includegraphics[angle=0,width=12cm]{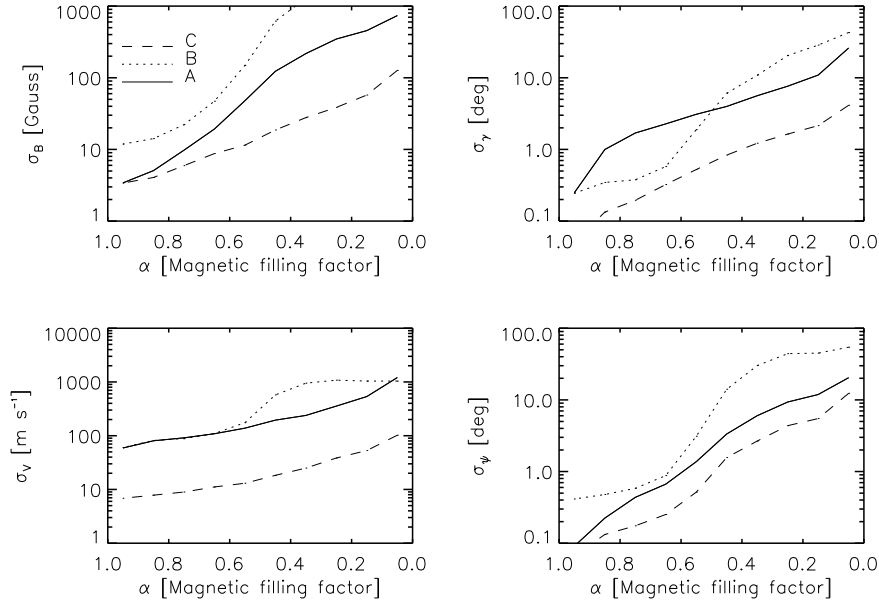}
\end{center}
\caption{Same as in Figure 6 but as a function of the filling
factor of the magnetic component ($\alpha$).}
\end{figure}

\section{Effect of the Photon Noise}%

The simulations in Section 5 did not include photon noise. Therefore, 
the next step in our analysis is to consider the simultaneous effect of the solar
{\it p} modes and the photon noise. To this end we have carried out two new
simulations where HMI profiles were obtained using modulation schemes A and C
and taking into account the expected photon noise (see Section 3.1). Errors
as a function of the filling factor of the magnetic component ($\alpha$) are presented 
in Figure 8 in thick lines (solid for scheme A and dashed for scheme C).
For comparison purposes we plot also the results from schemes A and C from Section 5
 (where only solar {\it p} modes were considered) in thin lines.
By comparing thin and thick lines in Figure 8 one can see that errors introduced
by the photon noise are dominant over those induced by the solar {\it p} modes.

An interesting effect appears after introducing the photon noise. 
As already mentioned in Section 5, the errors (when only {\it p} modes are considered) 
decrease with decreasing filling factor. This is generally also the case when
{\it p} modes and photon noise are considered simultaneously. However, for large filling
factors, $\alpha \rightarrow 1$, this effect is less pronounced. For some
physical parameters, like the field strength (Figure 8; top left panel)
errors even tend to slightly increase. This might be related 
to the fact that in our simulations, those points of the database 
with large filling factors are intended to represent
umbral points. The reduced light level in those regions translates
into a smaller signal-to-noise in the observed profiles, thereby 
increasing the errors in the retrieval of the magnetic properties.

\begin{figure}
\begin{center}
\includegraphics[angle=0,width=12cm]{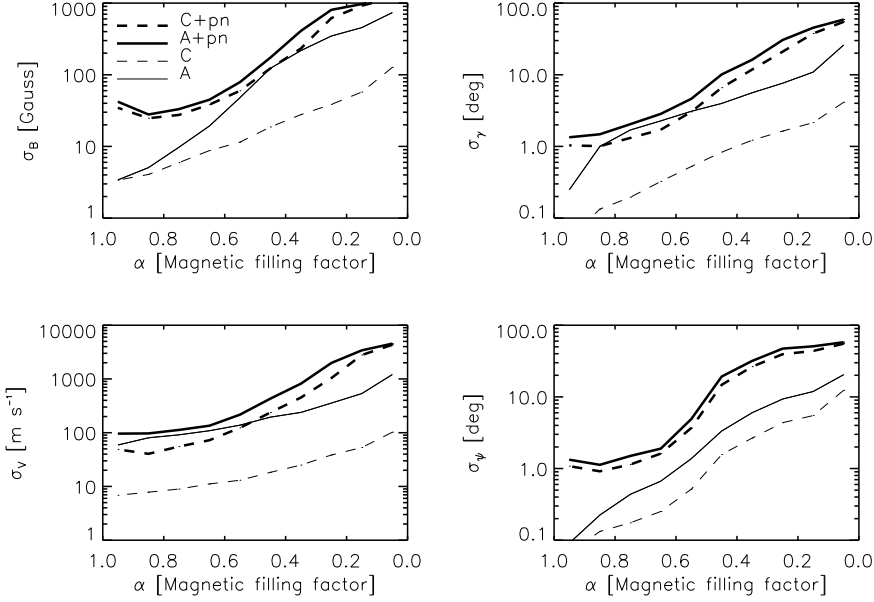}
\caption{Errors in the retrieved physical parameters as a function of the
filling factor of the non-magnetic component. Thin lines correspond to 
the results when only {\it p} modes are considered (same as in Figure 7). Thick
lines corresponds to results when solar {\it p} modes and photon noise (pn)
are taken into account. Schemes A and C are indicated in solid and 
dashed lines, respectively.}
\end{center}
\end{figure}

\section{Temporal Averages}%

In Section 3 we pointed out that at a given wavelength the crosstalk
induced by the solar {\it p} modes changes sign as the oscillation
goes from the increasing velocity regime to the decreasing part
of the wave (see Figure 3). By averaging in time the observed
signal, the {\it p}-mode crosstalk should be minimized. This effect is
shown in Figure 9 where we plot in circles the original
profiles as if no {\it p} modes existed: $\widetilde{\ve{I}}\obs(\lambda_i)$.
Overploted in dashed lines are Stokes $V$ observed during
consecutive time intervals: $[t_0,t_0 + 4PM]$, $[t_0 + 4PM,t_0 + 
8PM]$, {\it etc}. They are computed considering the {\it p} modes.
These dashed profiles recorded during $PM\Delta t$ seconds will
be referred to as {\it building block profiles}.
Since they are consecutive in time, the time-dependent 
behavior of the solar oscillation (see Figure 2; top panel) is sampled,
each of them contributing differently to the induced crosstalk. When all 
building blocks are averaged, the resulting profile (solid lines in Figure 9) 
closely resembles that without {\it p}-mode contribution (circles).

Figure 9 is obtained from a total averaging time of 480 seconds
(roughly 1.5 oscillatory periods). Note that, as imposed 
by Equation (6) and Equation (8), for modulation scheme A (left panel) the 
individual observations (dashed lines) are much more affected by 
the solar {\it p} modes than for scheme C (right panel). 
This does nothing but to stress the results from Figure 4.
Interestingly, the averaged Stokes $V$ (solid line) are in both
cases very similar. The reason for this is that during 480 seconds,
scheme A is able to observe Stokes $V$ as many as six times: $P=4
\rightarrow \widetilde{t} = 80$ seconds. Scheme C is able
to record Stokes $V$ only four times: $P=6 \rightarrow \widetilde{t} = 120$ 
seconds. This allows scheme A to provide a better sampling of the
solar {\it p}-mode oscillations, thus reducing more efficiently
the induced crosstalk.

Therefore, it is important to quantify how results compare 
when using different averaging times. In particular, whether 
scheme A becomes at some point more reliable than scheme C. 
To this end we have carried out several simulations using different averaging times with modulation 
schemes A and C. In these simulations photon noise was added 
to the building block profiles according to Section 3.1. The averaged profiles obviously
present a larger signal-to-noise. Some important properties of the simulations 
are presented in Table~VI.

\begin{table}
\begin{center}
\caption{Main properties of the simulations carried out in Section 6. First column
indicates the modulation scheme used. The asterisk indicates a
simulation done with the minimum time required to sample the full Stokes
vector for that particular scheme: {\it building block profiles}. Second column refers to the 
amount of total time used to average profiles. Note that it is always 
an integer multiple of the building block. Third column is the total
number of times that the full Stokes vector $\ve{I}$ is sampled during the time
indicated in the second column. Last column shows the noise level ($\sigma_i$)
in the averaged polarization profiles. This is indicated in units of
noise level in the building block observation using scheme C.}
\begin{tabular}{|c|c|c|c|}
\hline
Scheme & Averaging time [sec] & \# Stokes $\ve{I}$ & $\sigma_i$ \\
\hline
\hline
A & 880 & 10 & 0.369 \\
A & 560 &  7 & 0.462 \\
A & 240 &  3 & 0.707 \\
A$^{*}$ &  80 & 1 & 1.22 \\
\hline
C & 840 & 7 & 0.377 \\
C & 600 & 5 & 0.447 \\
C & 240 & 2 & 0.707 \\
C$^{*}$ & 120 & 1 & 1 \\
\hline
\end{tabular}
\end{center}
\end{table}

Figure 10 shows the errors in the physical parameters as a function of the 
time used to average. Note that the temporal evolution of solar
structures over the averaging time is neglected (except for the solar {\it p} modes).
In this sense the retrieved physical parameters should be considered
as an average of the physical conditions present in the solar atmosphere
during the time the data is acquired.
Results are presented considering three different solar 
structures: sunspots, plages and network regions. Those
definitions are given according to the filling factor
of the magnetic atmosphere. Sunspots: $\alpha \in [0.7,1]$,
 Plage: $\alpha \in [0.4,0.6]$, Network: $\alpha \in [0.1,0.3]$.
With the exception of the velocity, scheme A (thin lines) becomes more reliable
than scheme C (thick) using an averaging time as short as 240 seconds.
For large averaging times, both observing schemes become almost identical.
The HMI scientific goal is to provide the full magnetic field vector
with a cadence of ten minutes. At this level the errors 
for sunspot regions could be as little as $\sigma_B \sim 4$ Gauss, 
$\sigma_\gamma \sim 0.2^{\circ}$, $\sigma_\psi \sim 0.2^{\circ}$ and 
$\sigma_v \sim 20$ m s$^{-1}$ (see Tables VII and VIII for details).

\begin{figure}
\begin{center}
\begin{tabular}{cc}
\includegraphics[angle=0,width=5.5cm]{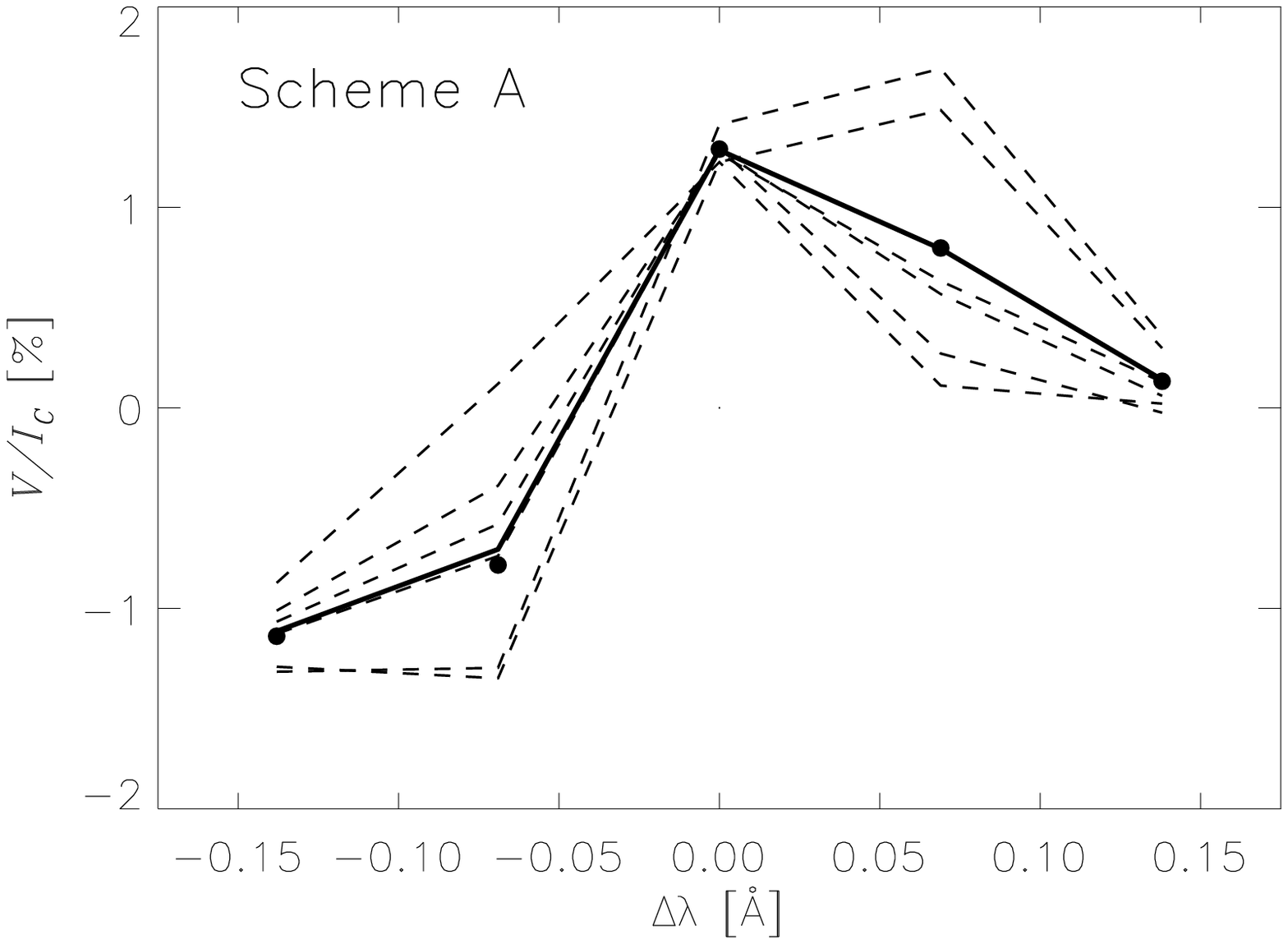} &
\includegraphics[angle=0,width=5.5cm]{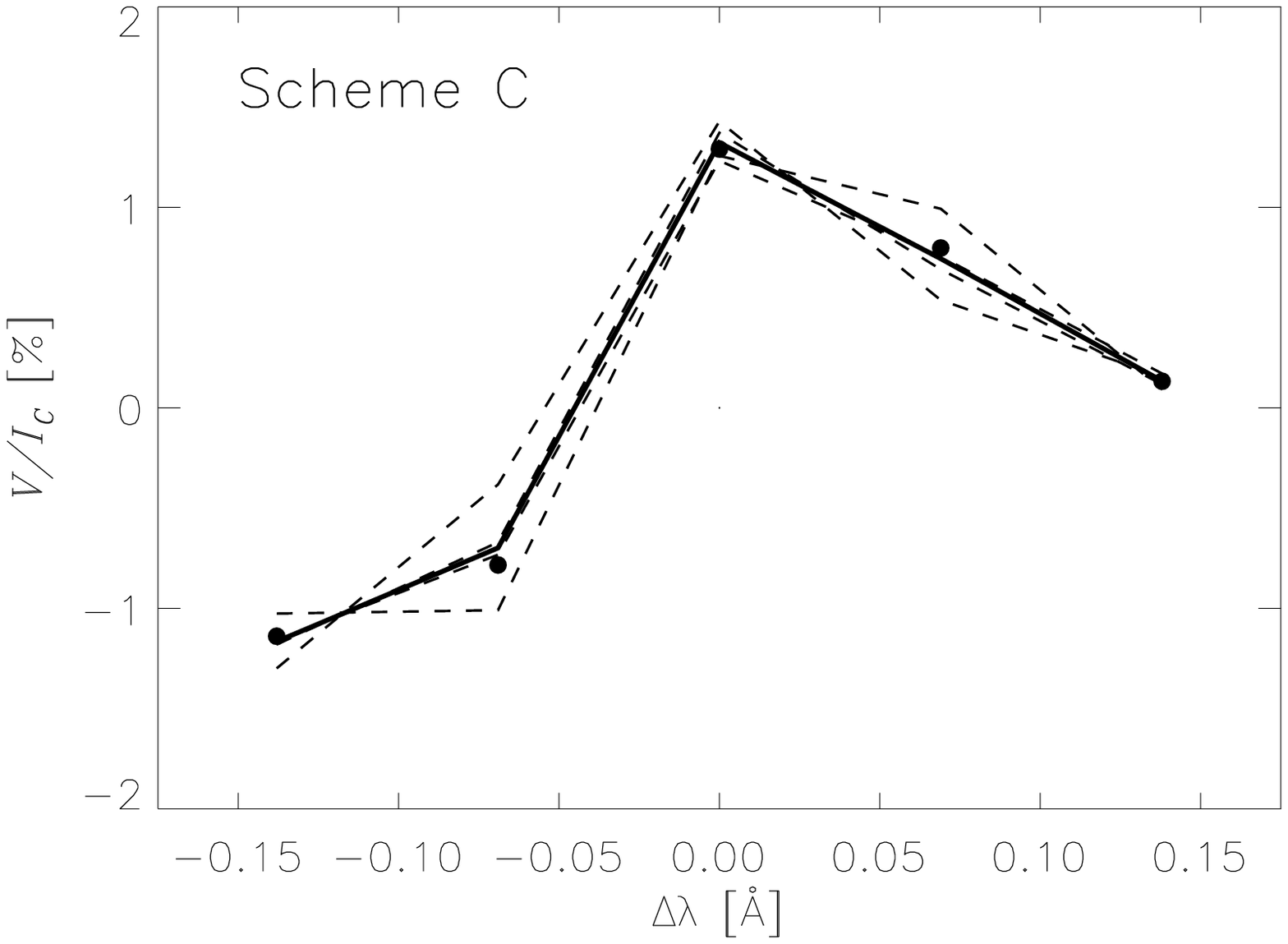}
\end{tabular}
\end{center}
\caption{{\it Left panel}: example of Stokes $V$ profiles
as observed by HMI neglecting the effect of the solar {\it p} modes (circles).
Dashed lines represent individual observations in consecutive time intervals
where solar {\it p} modes were accounted for using scheme A. Each observation requires 
$5P\Delta t$ seconds. The average of all of these observations is represented
by the solid line. {\it Right panel}: same but using scheme C. Photon noise
was not included in these examples.}
\end{figure}

\begin{figure}
\begin{center}
\includegraphics[angle=0,width=12cm]{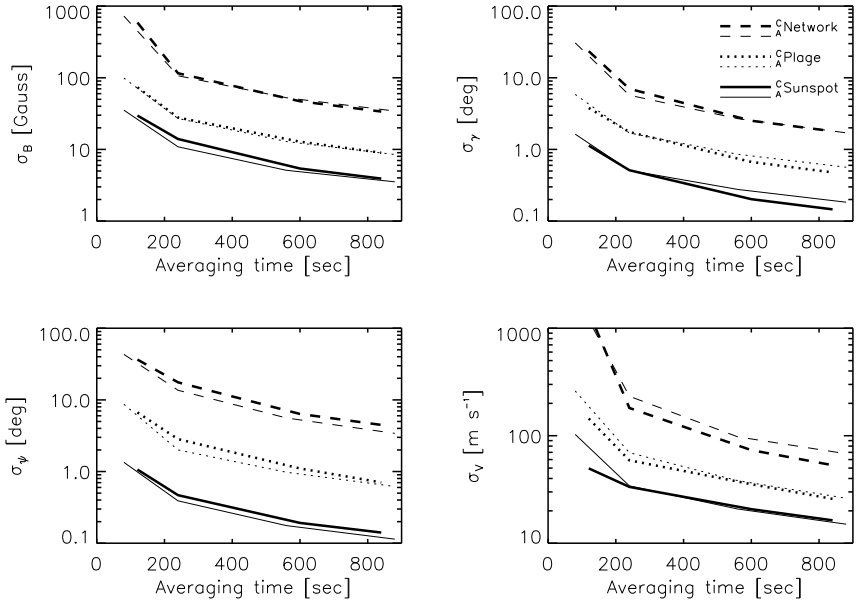}
\end{center}
\caption{Errors in the determination of the different physical parameters 
as a function of the averaging time. Thin lines indicate results for
modulation scheme A whereas thick ones are for scheme C. Different line
styles correspond to different observed solar regions: sunspots (solid),
 plage (dotted) and network regions (dashed).}
\end{figure}

\begin{table}
\begin{center}
\tabcolsep .6em
\begin{tabular}{|lcccc|}
\hline
Region & $\sigma_B$ & $\sigma_\gamma$ & $\sigma_\psi$ & $\sigma_{v}$\\
       & [Gauss] & [deg] & [deg] & [m s$^{-1}$]\\
\hline
\hline
Sunspot & 4 - 5 & 0.2 & 0.2 & 20\\
Plage & 10 & 1 & 1 & 30\\
Network & 40 - 50 & 2 - 3 & 5 - 7 & 60 - 90\\
\hline
\end{tabular}
\end{center}
\caption{Approximate errors in the determination of the properties (magnetic field
vector, velocity) of the magnetic component. This corresponds to a ten minute temporally averaged
data product.}
\end{table}
\begin{table}
\begin{center}
\tabcolsep .6em
\begin{tabular}{|lcccc|}
\hline
Region & $\sigma_B$ & $\sigma_\gamma$ & $\sigma_\psi$ & $\sigma_{v}$\\
       & [Gauss] & [deg] & [deg] & [m s$^{-1}$]\\
\hline
\hline
Sunspot & 30 & 1 & 1 & 40\\
Plage  & 80 - 100 & 4 & 6 - 8 & 100 - 200\\
Network & 400 - 700 & 30 & 30 - 50 & $>$ 1000 \\
\hline
\end{tabular}
\end{center}
\caption{Approximate errors in the determination of the properties (magnetic field
vector, velocity) of the magnetic component. This corresponds to a 120 second data product
(building block profiles) using Mod C.}
\end{table}

\section{Conclusions}%

The HMI instrument intends to provide accurate and continuous full-disk measurements 
of the solar magnetic field vector in the photosphere. We have identified and 
carefully studied the main sources of errors in HMI: limited spectral resolution,
 solar {\it p} modes (that affect our data as a consequence of the non-simultaneity in
the polarization measurements\footnote{Note that non-simultaneous observations will
not be affected by seeing induced cross talk since HMI is a space-borne instrument}) 
and photon noise. The most critical source of error
turns out to be the photon noise. Increasing the signal-to-noise ratio of the observations
is achieved with longer integration times. This is not trivial.
Increasing the exposure time per measurement may lead to saturation effects as we reach the limits
of the CCD's linear behavior. The only possibility left is to combine several 
measurements. We find that averaging data over ten minutes increases the signal-to-noise 
by a factor of two to three (see Table~VI) plus effectively removes the effect of the solar {\it p} modes.
This leads to very precise determinations of the time-averaged magnetic field vector, with an accuracy
better than five Gauss in field strength and 0.2 degrees in inclination and azimuth.

A faster and better correction for the {\it p} modes can be achieved in different ways. On the one hand,
the observations can be averaged over time using more sophisticated methods. In Section 7 a simple (unweighted)
average was used.  More sophisticated ways ({\it e.g.}: using appropriate weights) would lead to a faster
smearing of the solar {\it p} modes. On the other hand, it is possible to
interpolate the data so that we combine polarization measurements taken at the same time (see Equation (14)). 
This might correct the effects of the solar {\it p} modes even without time averaging. A last possibility
would be to apply a subsonic filter to the raw ({\it i.e.}: monochormatic and not demodulated) data (Rutten,
 Wijn, and S\"utterlin, 2004). Although these possibilities are  worth investigating, they will not significantly change 
our results since the main source of errors is the photon noise.

A critical point we have not mentioned concerns the contribution of
the non-magnetic atmosphere $\ve{I}\nm$ (see Equation (10)). In all our simulations we have considered
$\ve{I}\nm$ as known (with the exception of the filling factor $\alpha$). This is a best case
scenario. In reality, the $\ve{I}\nm$ contribution can arise 
because of scattered light from the optical elements of the instrument, or due to the presence of
a real non-magnetic atmospheric component within the resolution element 
of the observations.

In the first case (instrumental scattered light), it might be possible to
characterize $\ve{I}\nm$ with laboratory tests. 
In the second case, its origin would be solar, and therefore we should have considered
new free parameters in our inversion to describe $\ve{I}\nm$, such as: ${\rm v^{\rm LOS}_{\rm nm}}, 
{\rm S_{\rm nm}}, {\rm \dot{S}_{\rm nm}}$. We have neglected this in order to keep the number of free
parameters (listed in Table~IV) within reasonable limits and below the number of data points (20).
A necessary extension of this work should address this point carefully by both {\it (a)} 
including the new free parameters and, {\it (b)} by using a different $\ve{I}\nm$ 
in the synthesis and the inversion so that we can quantify the errors introduced by
a poorly known $\ve{I}\nm$. Obviously the larger $\alpha$, the smaller these errors
will be. Therefore, for sunspot regions (Figure 10 and Tables~VII and VIII) the results presented
in this work should be robust.

Another very important issue concerns the speed of the inversion process (see Section 4). 
The employed method is a traditional non-linear least-squares fitting algorithm. This method will
not ultimately be used for the inversion of HMI data. Our simulations with the analysis of 20 000 profiles
took an average of 24 hours to complete. HMI requires the processing of millions of profiles
per minute. Even those methods, considered to be fast, such as Principal Component Analysis 
(Rees {\it et al.}, 2000) or Artificial Neural Networks (Socas-Navarro, 2003), will be
challenged by HMI requirements. This will be the subject of a future investigation.

\section*{References}

\begin{itemize}
\item Cabrera Solana, D., Bellot Rubio, L.R, and del Toro Iniesta, J.C.: 2005, {\it Astron. Astrophys.} , {\bf 439}, 687.
\item Graham, J.D., L\'opez Ariste, A., Socas-Navarro, H., and Tomczyk, S.: 2002, {\it Solar Phys.}, {\bf 208}, 211.
\item Graham, J.D., Norton, A.A, Ulrich, R.K. et al.: 2003, HMI-TN-03-002.
{\fontfamily{pcr}\selectfont http://hmi.stanford.edu/doc/Tech\_Notes/TechNoteIndex.html}
\item Jefferies, J.T. and Mickey, D.L.: 1991, {\it Astrophys. J.}, {\bf 372}, 694.
\item Landi Degl'Innocenti, E.: 1991, in F. S\'anchez, M. Collados and M. V\'azques (eds.),
{\it Solar Observations: Techniques and Interpretation. First
Canary Islands Winter School of Astrophysics.}. Cambridge University Press.
\item Norton, A.A.: 2000, Ph.D Thesis, Univ. California Los Angeles.
\item Norton, A.A., Pietarila Graham, J., Ulrich, J.K., Schou, J., Tomczyk, S., Liu, Y., Lites, B., 
Lopez Ariste, A., Bush, R., Socas Navarro, H., Scherrer, P., 2006, {\it Solar Phys.}, in press. 
\item Press, W.H., Flannery, B.P., Teukolsky, S.A., and Vetterling, W.T.: 1986, 
Numerical Recipes, Cambridge University Press.
\item Rees, D.E., L\'opez Ariste A., Thatcher, J., and Semel, M.: 2000, {\it Astron. Astrophys.}, {\bf 355}, 759.
\item Ruiz Cobo, B. and del Toro Iniesta, J.C.: 1992, {\it Astrophys. J.}, {\bf 398}, 375.
\item Rutten, R.J., de Wijn, A.G. \& S\"utterlin, P.: 2004, {\it Astron. Astrophys.}, {\bf 416}, 333.
\item Scherrer, P.H., Bogart, R.S., Bush, R.I. et al.: 1995, {\it Solar Phys.}, {\bf 162}, 129.
\item Socas-Navarro, H.: 2003, {\it Neural Networks}, {\bf 16}, 355.
\item del Toro Iniesta, J.C. 2003, Introduction to Spectropolarimetry, Cambridge University Press.
\item del Toro Iniesta, J.C. and Collados, M.: 2000, {\it Applied Optics}, {\bf 39}, 1637.
\end{itemize}

\end{document}